\documentclass[aps,prl,a4paper,superscriptaddress,twocolumn,showpacs]{revtex4} 
\usepackage{graphics}
\usepackage{setspace}
\begin{document}
\title{Strong Correlations in Electron Doped Phthalocyanine Conductors Near Half Filling}

\author{Erio Tosatti}
\altaffiliation[]{tosatti@sissa.it}
\affiliation{International School for Advanced Studies (SISSA), and INFM
Democritos National Simulation Center, Via Beirut 2-4, I-34014 Trieste, 
Italy}
\affiliation{International Centre for Theoretical Physics
(ICTP), P.O.Box 586, I-34014 Trieste, Italy}

\author{Michele Fabrizio}
\affiliation{International School for Advanced Studies (SISSA), Via Beirut 2-4, 
I-34014 Trieste,Italy}
\affiliation{International Centre for Theoretical Physics
(ICTP), P.O.Box 586, I-34014 Trieste, Italy}

\author{Jaroslav T\'{o}bik}
\affiliation{International School for Advanced Studies (SISSA), and INFM
Democritos National Simulation Center, Via Beirut 2-4, I-34014 Trieste, 
Italy}

\author{Giuseppe E. Santoro}
\affiliation{International School for Advanced Studies (SISSA), and INFM
Democritos National Simulation Center, Via Beirut 2-4, I-34014 Trieste,
Italy}
\affiliation{International Centre for Theoretical Physics
(ICTP), P.O.Box 586, I-34014 Trieste, Italy}

\date{\today}
\begin{abstract}

We propose that electron doped nontransition metal-phthalocyanines (MPc) like ZnPc 
and MgPc, similar to those very recently reported, should constitute 
novel strongly correlated metals. Due to orbital 
degeneracy, Jahn-Teller coupling and Hund's rule exchange, and with 
a large on-site Coulomb repulsion, these molecular conductors should display, 
particularly near half filling at two electrons/molecule, very unconventional 
properties, including Mott insulators, strongly correlated superconductivity, 
and other intriguing phases.

\end{abstract}

\maketitle

Building novel metals by doping molecular crystals 
such as polyacetylene, fullerenes, TTF-TCNQ, (TMTSF)$_2$X, (TMTTF)$_2$X and 
(BEDT-TTF)$_2$X salts, etc., 
is a well trodden route \cite{review,forro}, but remains an exciting and ever moving front.
Very recently the Delft group showed that thin films of transition metal 
phthalocyanines (MPcs) FePc, CoPc, NiPc, CuPc, initially insulating, can be 
turned genuinely metallic through potassium doping \cite{craciun}.
Electron doping appears to take place largely in the twofold degenerate 
lowest unoccupied $e_g$ molecular orbital (LUMO) \cite{liao,note1} of the molecules. 
The simplest rigid band model naturally explains, according to
\cite{craciun}, why the MPcs, initially insulating when pristine 
($n=0$) \cite{note2}, become metallic upon increasing doping ($0 < n < 4$), 
ending up again as insulators at full doping ($n=4$). It is not yet clear 
whether stoichiometric compound phases may exist here as they do in alkali 
doped fullerides, but there is an otherwise striking overall analogy, to 
the point that even the conductance values reported at optimal metallic 
doping are close in magnitude and temperature (in)dependence to those of 
K$_x$C$_{60}$ films.

While the pursuit of metallicity-cum-magnetism in the transition metal
doped MPcs \cite{note1} will in the future constitute an interesting goal in 
itself, the scope of this note is to point out newer directions and 
possibilities  that can make slightly different metallic doped MPcs, 
yet to be realized, potentially even more exciting. Doped MPcs are not, 
we claim, regular metals, but constitute strongly correlated electron 
systems, akin to doped two-band Mott insulators. Characterized by a strong 
on-site electron repulsion, by the $e_g$ orbital degeneracy, and by 
intra-site Jahn-Teller and electron-electron multiplet couplings, the doped 
MPc molecular system should approach near $n = 2$, as foreshadowed by 
recent studies\cite{capone03,han03,DeLeo} a novel unstable fixed point 
heralding a wealth of possible low-temperature phenomena and phases.
We conduct in this letter a preliminary exploration of this scenario by addressing 
theoretically -- and thus proposing the experimental realization of -- new metals obtained 
through electron doping of ZnPc, MgPc and other such non-transition metal MPcs. 
The extra electrons should flow into the MPc 2$e_g$ lowest unoccupied molecular
orbital (LUMO)\cite{liao}, to form a two band metal. The LUMO states are ligand shell 
orbitals loosely surrounding the central metal ion, with a large 
intra-molecular Coulomb repulsion $U$ in comparison with 
the narrow electron bandwidth $W$ expected from weak intermolecular electron 
hopping.
In the neighborhood of half filling, $n=2$, 
a doubly degenerate $e_g$ orbital also has a molecular Hund's rule exchange 
$J$ (lowering the triplet state by 4$|J|$ relative to the singlet), 
and a Jahn-Teller (JT) coupling of the electronic state to 
$B_{1g}$ and $B_{2g}$
MPc molecular vibrations, lowering the singlet state energy by an amount 
$E_{JT}$. Thus there will be some cancellation between the 
two couplings. These ingredients are similar in nature to those 
present in fullerides, where orders of magnitude are $U=1$ eV, 
$E_{JT}= 0.15$ eV, 
$J=0.03$ eV, and $W = 0.5$ eV. Although the corresponding values in MPcs 
are not yet so accurately known, we obtain estimates that are surprisingly close.
The similarity noted in Ref.\ \cite{craciun} between alkali doped MPcs and 
fullerides\cite{forro} may thus reach much deeper than 
just the dopability and the closeness of mobilities pointed out experimentally. 
As in fullerides, the resulting phase diagram will depend upon the balance 
between low and high spin states of the molecular ion. Unlike fullerides,
however, the twofold degeneracy will reveal a potentially richer overall 
phase diagram.

The Hamiltonian for electrons in the $e_g$-orbitals of molecule $i$ is
\begin{eqnarray}
\hat{H}_{i,mol} &=& \frac{U}{2}\, \hat{n}_i^2 + 
\frac{J}{2}\, \left( \hat{\tau}_{i,x}^2 + \hat{\tau}_{y,i}^2\right)\nonumber \\
&& +g\left(q_{i,1}\, \hat{\tau}_{i,x} + q_{i,2}\,\hat{\tau}_{i,y}\right)
\nonumber \\
&&+ \frac{\hbar\omega_0}{2} \left( q_{i,1}^2 + p_{i,1}^2 + q_{i,2}^2 + 
p_{i,2}^2\right) \;,
\label{Ham}
\end{eqnarray}
where $\hat{n}_i$ is the overall occupation at site $i$,
$\hat{\tau}_{i,\alpha}$, with $\alpha=x,y,z$, are pseudo-spin operators 
spanning the twofold degeneracy 
$
\hat{\tau}_{i,\alpha} = \sum_{a,b=1}^2 \, \sum_{\sigma} \,
c^\dagger_{i,a\sigma}\, \left(\tau^{\alpha}\right)_{ab}\, 
c^{\phantom{\dagger}}_{i,b\sigma} \;,
$
where $\tau^{\alpha}$ are Pauli matrices. Here
$c^\dagger_{i,a\sigma}$ denotes creation 
of electrons with spin $\sigma$ at site $i$ in orbital $a=1,2$ of the LUMO $e_g$ doublet.
The last two terms in (\ref{Ham}) describe JT coupling of
strength $g$ to a doubly degenerate 
vibration with frequency 
$\hbar\omega_0$, coordinates $q_{i,1}$ and $q_{i,2}$ and 
momenta $p_{i,1}$ and $p_{i,2}$, where $E_{JT}=8g^2/\hbar \omega_0$. 
A single but degenerate mode is assumed as a simplifying 
approximation, replacing the actual 14 pairs of nondegenerate JT
modes $B_{1g}$ and $B_{2g}$ allowed by the MPc $D_{4h}$ symmetry.
As detailed in note~\cite{parameters} an estimate of parameters appropriate to 
MPcs such as ZnPc, MgPc indicates $U \sim 1 eV$ and, for $n=2$, a singlet 
energy gain $E_{JT} - 4|J| \sim$ 0.06-0.07 $eV$ over Hund's rule triplet.
This prevalence of singlet over triplet, the main ingredient potentially 
leading to s-wave superconductivity, could be mitigated or even reversed in a solid 
state compound; we shall therefore examine both possibilities 
in the rest of this paper.

In the MPc molecular crystal, electrons can quantum 
mechanically hop between neighboring sites 
\begin{equation}
\hat{H}_{hop} = \sum_{ij}\sum_{a,b=1}^2\sum_\sigma \, t_{ij}^{ab}\,
c^\dagger_{i,a\sigma}\, c^{\phantom{\dagger}}_{i,b\sigma} \;.
\label{H-hop}
\end{equation}
Assuming first a generic three-dimensional structure with $\nu$ neighboring molecules 
and a weak isotropic first-neighbor hopping $t$, the LUMO $e_g$ orbitals will give rise to 
a pair of narrow molecular bands of width $W=2|t|\nu$. (We will discuss the additional 
signature of a quasi-one-dimensional structure of these compounds in the 
final part of the paper.) Before doping we generally expect 
a narrow band $W \sim 0.3$ eV (see below), due to very weak van der Waals 
bonding between molecules. While $W$ will eventually depend on the intermolecular 
geometry, including the way that may be altered by doping, we will 
assume $W$ to remain of approximately the same magnitude after doping.
In addition to (\ref{Ham}) and (\ref{H-hop}) we should in principle 
include additional terms such as the dispersion of vibrations (to account 
for the coupling between the JT distortion of a molecule 
and those of other molecules), inter-molecular electron-electron
interactions, and more. We will presently neglect them, concentrating 
on the basic Hamiltonian
\begin{equation} \label{totalH}
\hat{H} = \hat{H}_{hop} + \sum_i \, \hat{H}_{i,mol} \;.
\label{Ham-final}
\end{equation}
Because $U/W$ is large, and for $n \sim 2$, this system is generally close to 
a Mott transition. There, the quasiparticle band is very narrow and it is a good
approximation to neglect retardation, replacing (\ref{Ham}) with 
\begin{equation}
\hat{H}_{i,mol} \rightarrow \frac{U}{2}\, \hat{n}_i^2 +
\frac{J_{eff}}{2}\, \left( \hat{\tau}_{i,x}^2 + \hat{\tau}_{y,i}^2\right) \;,
\end{equation}
where $J_{eff}=J-E_{JT}/4$.

The further simplifying assumption of nearest-neighbor hopping that are
diagonal in the orbital indices \cite{note3} brings this Hamiltonian to coincide 
with that recently studied by Dynamical Mean Field Theory (DMFT)\cite{DMFT} in 
Refs.\ \cite{capone03,han03}. Close to $n=2$ it was shown to display 
a rich phase diagram 
including, besides the regular metal, a Mott insulator, a strongly
correlated superconductor, and a pseudo-gap metal.
Within DMFT, the lattice model (\ref{totalH}) 
is mapped onto an Anderson impurity self-consistently coupled to a conduction 
electron bath \cite{DMFT}. The many-body physics of the 
Anderson impurity and its fixed points foreshadow that of the infinite 
many-body system\cite{DeLeo}, as confirmed by the calculated DMFT phase
diagram~\cite{capone03,han03}. 
We exploit here these results to analyse the implications for the phase 
diagram of half-filling doped MPcs considering both possibilities, 
namely: ({\sl i}) singlet ground state or ({\sl ii}) triplet ground state. 

({\sl i}) $J_{eff}<0$ -- The $n$=2 molecular ground state is a 
non degenerate spin singlet accompanied by a dynamical Jahn-Teller effect. 
If in the solid $U/W$ is larger than the critical Mott value 
(between 1 and 2, depending in detail on J and $E_{JT}$) the half filled MPc 
will realize a non-magnetic, singlet Mott insulator \cite{fabrizio97}. 
For $U$ just below that value, or for light doping away from $n$=2, 
the MPc solid will be lightly metallic. In this ``doped Mott insulator'' regime
the Anderson impurity displays instabilities against symmetry-broken phases 
in both particle-particle and particle-hole channels; which instability
will eventually win out depends on band-structure 
and coupling details. If nesting or other band structure singularities 
are absent or very weak, the particle-particle instability should dominate 
leading to an $s$-wave superconductor with order parameter
\begin{equation}
\Delta_{SC} = \langle c^\dagger_{1\uparrow}c^\dagger_{2\downarrow}
+ c^\dagger_{2\uparrow}c^\dagger_{1\downarrow}\rangle \not = 0 \;.
\label{Delta-SC}
\end{equation}
This state is a strongly correlated superconductor (SCS)
of a kind first pointed out in Ref.\ \cite{Capone-Science}, further
confirmed in the two band model in Refs.\ \cite{capone03,han03}. 
It was also suggested\cite{Capone-Science,capone03} -- although without a calculation of $T_c$ -- 
that SCS is a ``high temperature superconductor'' in the sense that
the SCS superconducting gap may reach values several orders of magnitudes 
greater than the corresponding BCS value (that would be attained for $U=0$),
$\Delta_{BCS} \sim \hbar \omega_0\, {\rm e}^{-1/\lambda} $,
where $\lambda = 2\rho_0 \, |J_{eff}|$, $\rho_0$ is the density of states
at the Fermi level per spin and band.

If, on the contrary, the particle-hole instability channel was favored by some 
detail of the band structure such as Fermi surface nesting, then one of two 
alternative symmetry broken phases can be expected. 
The first is a trivial one, with order parameter 
\begin{equation}
\Delta_{JT} = \cos\phi \, \langle \hat{\tau}_x \rangle
+ \sin\phi \, \langle \hat{\tau}_y \rangle \not = 0 \;,
\end{equation}
corresponding to a cooperative Jahn-Teller distorted state, either modulated or 
uniform, the angle $\phi$ reflecting U(1) orbital symmetry breaking. The
model in fact possesses O(2) orbital symmetry, which naturally decomposes into
Z$_2~\otimes$ U(1). Z$_2$ is the discrete symmetry $\tau_z ~\to~ -\tau_z$, 
while U(1) represents invariance under rotations around the pseudo-spin $z$-axis. 
As a consequence of the static cooperative Jahn-Teller distortion, a conventional 
insulating band gap or at least a strong lowering of metallic density of states 
should take place at the Fermi level. More interesting is the alternative instability, 
associated with a different order parameter in the particle-hole channel
\begin{equation}
\Delta_{PT} = \sum_{\alpha\beta}
\langle c^\dagger_{1\alpha}\, \vec{\sigma}_{\alpha\beta}\cdot \vec{S}\,
c^{\phantom{\dagger}}_{1\beta}\rangle -
\langle c^\dagger_{2\alpha}\, \vec{\sigma}_{\alpha\beta}\cdot \vec{S}\,
c^{\phantom{\dagger}}_{2\beta}\rangle\not = 0 \;,
\label{PT}
\end{equation}
breaking both orbital Z$_2$ and spin SU(2) symmetry, $\vec{S}$ being the 
direction along which the spin SU(2) symmetry is broken. We note that the 
LUMO $e_g$ orbitals are mainly localized onto the MPc 
four N-atom ring and have odd parity with respect to the central metal ion. 
Therefore the discrete Z$_2$ symmetry translates into parity in MPcs, 
whence $\Delta_{PT}\not = 0$ represents a spin current flowing 
on the Pc ring, a very intriguing possibility. 
This kind of phase should be either insulating or a poor metal like the 
cooperative JT phase. Unlike the latter, it should break both time 
reversal and parity, conserving only the product of the two. These properties are 
reminiscent of a phase discussed some time ago by Varma \cite{varma}.

({\sl ii}) $J_{eff}>0$ -- In this case (less likely in MPcs 
than (i), as said above) the isolated $n$ =2 molecular ion is a spin 
triplet. The strongest instability here is towards bulk magnetism\cite{DeLeo}, 
ferro or antiferro depending on the band structure, a state which 
may or may not be accompanied by parity-symmetry breaking or by a Jahn-Teller 
distortion. If (depending on lattice structure) magnetism was sufficiently frustrated, 
then it would be possible for spin-triplet superconductivity to appear with order parameter
\begin{equation}
\Delta_{SC} = \langle c^\dagger_{1\uparrow}c^\dagger_{2\downarrow}
- c^\dagger_{2\uparrow}c^\dagger_{1\downarrow}\rangle \not = 0 \;.
\label{Delta-TSC}
\end{equation}
Since orbitals 1 and 2 have opposite parity, this translates into a
$p$-wave spin-triplet superconducting order parameter. This is an
interesting possibility since, at variance with conventional $p$-wave 
superconductors, the pairing function is nonzero on the molecule.
According to the DMFT analysis of Ref.\ \cite{han03}, where this 
kind of triplet suoerconductor was first discussed, this superconducting 
instability is also enhanced by the proximity of a Mott insulating phase, 
although not as dramatically as in case ({\sl i}). 

\begin{figure}[t]
{\par\centering \resizebox*{!}{7cm}{\rotatebox{0}{\includegraphics{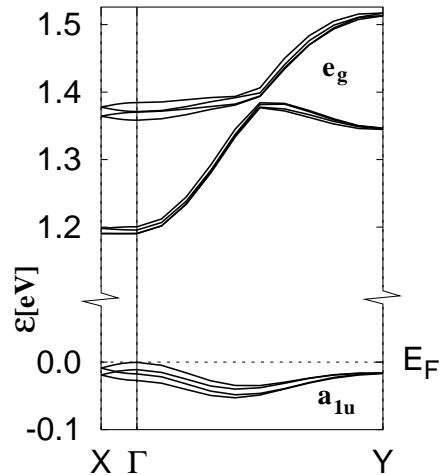}}}\par}
\label{fig-bands}
\caption{Density functional electronic band structure of undoped
$\alpha$ structure MgPc. The molecular stack direction is $\Gamma - Y$.
The crystal structure was taken from Ref.\ \cite{Brown}. The monoclinic 
cell ($a=26.29$, $b=3.818$, $c=23.92$\AA, $\beta=94.6$) contains four
molecules and has symmetry $C2/n$.}
\end{figure}

Much of the uncertainty in the above scenario depends on our ignorance
of the doped MPc film crystal structure, so far treated generically.
A second guess based on the crystal structure of pristine MPcs  
suggests that the alkali doped crystals could, similar to the undoped alpha-phase
of the CuPc films\cite{craciun}, be made up of quasi-one-dimensional (1D) chains. We carried 
out density functional electronic structure calculations of MgPc assuming 
an alpha phase structure with $n$ = 0\cite{jaro}, and found an 
intra-chain $e_g$ bandwidth $W\sim $ 0.3 $eV$ but an inter-chain bandwidth 
between one and two orders of magnitude narrower (see Fig.\ \ref{fig-bands}). If doped MPcs were
indeed this close to 1D, then our earlier DMFT-based analysis, valid for the 
opposite limit of large coordination lattices, should
be replaced by another where the additional effect of quantum fluctuations 
and other 1D specific anomalies are properly treated.

We studied this 1D model by adapting the two-loop renormalization 
group (RG) equations derived by Ref.\ \cite{michele} to the two-band model. 
The fixed point Hamiltonian towards which the model flows under RG was 
analyzed by means of bosonization\cite{michele,shura}. 
At strict half-filling we found a spin-gapped insulator for any value of $U$. 
However, for large $U$, this was a spin-liquid Mott insulator, {\it i.e.} 
a spin-gapped phase without any symmetry breaking. That corresponds
either to the DMFT singlet Mott state found with $J_{eff}<0$, 
or to a Haldane spin-1 chain for $J_{eff}>0$. Here, the DMFT metallic 
phase at small $U$  is replaced by a spontaneously dimerized insulator, 
{\it i.e.} a spin-gapped phase with broken translational symmetry, 
for either sign of $J_{eff}$. Away from half-filling the model turns 
metallic: nonetheless we find that the spin-gap survives. In particular, 
in the large $U$ regime likely pertinent to the MPcs,  
the spin-liquid Mott insulator at $J_{eff}>0$ turns at low doping into a spin-gapped metal 
with dominant fluctuations in the SC singlet channel (\ref{Delta-SC}) 
and sub-dominant fluctuations in the particle-hole 
$4k_F$ channels identified by Ref.\ \cite{DeLeo}. The weakly-doped spin-1 chain 
at $J_{eff}<0$ feels on the contrary more the effects of the reduced dimensionality which 
causes the spin gap to survive in the metallic phase, hence preventing long-range  
spin-triplet superconducting correlations. Indeed, in 1D the leading SC fluctuations 
appear in a spin-singlet orbital-singlet but space-odd particle-particle channel\cite{M&S}.

Weak but nonzero interchain interactions will in a hypothetical doped MPc 
with alpha phase-like structure turn all these instabilities into properly 
long-range ordered phases at sufficiently low temperatures. While our present state 
of ignorance prevents further in depth discussion, the expected emerging scenario 
is parallel to that obtained by DMFT, apart from those differences specific to 1D, 
like dimerization at $n=2$ or the persistence of a spin gap away from half-filling 
for regular Hund's rules.  

Summarizing, we conclude first of all that stoichiometrically doped nontransition 
MPcs should display Mott insulating phases at all integer fillings, in particular 
at half filling. Analysis of a simple model (\ref{Ham}) revealed a stunning variety 
of phases in the immediate neighborhood of the Mott metal-insulator transition 
near half filling. 
The variety is greater than either that predicted for the three band  
case fullerides\cite{Capone-Science}, or
that experimentally known in fullerides\cite{forro} as well as in doped organics 
(with lower symmetry than MPcs) near half filling\cite{review}.

Many of the phases described above will individually merit an in-depth study. 
At the theoretical level, approximations will need to be improved, for 
example by treating properly the non-degeneracy of the $B_{1g}$ and $B_{2g}$ 
molecular vibrations, the non-diagonal hopping, the dispersion of vibrations, 
etc. At the moment, nonetheless, it seems fair to say that more direct 
experimental input is urgent and essential. It will be important to dope and 
metallize non-transition MPcs, and to investigate whether they can be made 
at least locally stoichiometric. Subsequent examination by, {\it e.g.}, 
scanning tunneling spectroscopy could search for the Mott insulating state 
and begin broaching in its neighborhood this fantastic scenario.

Work in SISSA/ICTP/Democritos was sponsored by MIUR FIRB RBAU017S8R004,
FIRB RBAU01LX5H, and MIUR COFIN 2003, as well as by INFM. We acknowledge
illuminating discussions with A. Morpurgo, A. Nersesyan, and N. Manini.


\begin{thebibliography}{30}
\bibitem{note1} 
In most MPcs described in Ref.\ \cite{craciun},
there is a partly filled $d$-level. Strongly localized and with a 
large on-site Coulomb repulsion, that constitutes a nontrivial magnetic complication. 
The potentially interesting interplay of metallic conduction in the LUMO 
with these localized $d$ electrons is beyond the scope of this work.
\bibitem{note2} 
MnPc, FePc, CoPc and CuPc are open shell molecules \cite{liao}, so that the 
pristine parent insulators must themselves constitute (single-band) Mott-Hubbard 
insulators.
\bibitem{parameters} An estimate of $U$ is given by
$U = (E_2-E_3) - (E_1-E_2)= A_1 - A_2$, where $E_n$ is the total 
energy of the molecular ion with $n$ electrons,
and $A_n$ its electron affinity. 
Calculations for isolated ZnPc(n-) and MgPc(n-) \cite{liao} find $A_1\sim 3.7$ eV
and $A_2\sim 0.7$ eV, whence $U \sim 3$ eV,
essentially identical to that of fullerene in vacuum\cite{manini}.
In crystalline MPcs the screening could be similar to that of fullerene \cite{lof}
suggesting $U = 1\div 1.5$ eV as a crude estimate.
Fits of the
multiplet spectra of ZnPc~\cite{minor} suggest that in the LUMO $e_g$ orbitals 
$4|J|$ could be $\approx 0.15-0.2$ eV. DFT calculations of ZnPc(2-) and MgPc(2-)
yield singlet ground states, with 
$E_{JT}-4|J| \sim$ 0.06 and 0.07 eV respectively\cite{liao}.
Resonance studies on ZnPc(2-) and MgPc(2-) in solution \cite{singlets}
confirms that these dianions could indeed be singlets.
\bibitem{note3}
While the diagonal hopping approximation is not particularly 
realistic in the present context, it does not change the nature of the problem, 
and provides useful physical understanding for application to the doped MPcs.
\end{thebibliography}
\end{document}